\documentclass[%
preprint,
 amsmath,amssymb,
 aps,
]{revtex4-2}
\usepackage{xcolor}
\usepackage{graphicx}
\usepackage{dcolumn}
\usepackage{bm}
\usepackage{amsmath,amssymb}

\begin{document}

\preprint{APS/123-QED}

\title{Stress overshoot, hysteresis and Bauschinger effect in sheared dense colloidal suspensions}

\author{Dmytro Kushnir}
\affiliation{IPCMS/CNRS, 23 rue du Loess 67034 Strasbourg, FRANCE
}%
\author{Céline Ruscher}%
\affiliation{%
 ICS/CNRS 23 rue du Loess 67034 Strasbourg, FRANCE
}%

\author{Eckhard Bartsch}
\affiliation{
Institut für Physikalische Chemie and Institut für Makromolekulare Chemie, Albert-Ludwigs-Universität, D-79104 Freiburg, GERMANY
}
\author{Fabrice Thalmann}
\affiliation{%
 ICS/ CNRS 23 rue du Loess 67034 Strasbourg, FRANCE
}%
\author{Pascal Hébraud}
\affiliation{IPCMS/ CNRS, 23 rue du Loess 67034 Strasbourg, FRANCE
}%

\date{\today}
\begin{abstract}
The mechanical non-linear response of dense Brownian suspensions of polymer gel particles is studied experimentally and by means of numerical simulations. It is shown that the response to the application of a constant shear rate depends on the previous history of the suspension. When the flow starts from a suspension at rest, it exhibits an elastic response followed by a stress overshoot and then a plastic flow regime. Conversely, after flow reversal, the stress overshoot does not occur, and the apparent elastic modulus is reduced while numerical simulations reveal that the anisotropy of the  local microstructure is delayed relative to the macroscopic stress.
\end{abstract}

\maketitle

\section{\label{sec:intro}Introduction}

Materials deform or flow under the application of stress. Many exhibit both behaviors and their response depends on the amplitude of the applied stress, and very often it depends also on the past history of the materials.  A common feature of crystalline materials and amorphous systems, such as foams, emulsions or concentrated suspensions is the existence of a transition between an elastic response at small deformation amplitudes to a plastic response, where the system flows, at larger amplitudes.  In general, under an increasing applied deformation, the stress first linearly increases, and reaches a steady state value at large deformations, but goes through a maximum value before it reaches its steady-state value. This stress overshoot is a common feature of many amorphous materials, such as emulsions~\cite{papenhuijzen1972}, foam~\cite{Khan_1988} suspensions~\cite{Nagase_1986} and its amplitude depends on the increase rate of the deformation~\cite{Dzuy_1983}.

Numerical simulations of binary Lennard-Jones glasses suggest that the overshoot value, $\sigma_{over}$, scales with the applied shear rate $\dot{\gamma}$ as $\sigma_{over} \propto \ln\dot{\gamma}$ ~\cite{Varnik_2004, Rottler_2005}. Conversely, numerical simulations of attractive spherical particules exhibit a power law behavior,  $\sigma\propto\dot{\gamma}^{0.5}$~\cite{Whittle_1997} and neutralized carbopol microgel suspensions exhibit a weaker dependency $\sigma_{over}\propto\dot{\gamma}^{0.13}$~\cite{Divoux_2011}. Moreover, experiments have shown that attractive colloidal suspensions exhibit two distinct yield points, due to the breaking of inter-cluster bonds and to collisions between clusters respectively~\cite{Koumakis_2011}. The second yield point is characterized by a stress overshoot whose value depends on the applied shear rate, as $\sigma_{over}\propto\dot{\gamma}^{0.5}$, the exponent being almost independent of the concentration.\\ 
The yielding mechanisms of hard sphere suspensions are \textit{a priori} different as they do not involve the breaking of attractive bonds between particules, but the escape from cages is the main mechanism of yielding. Yield is related to the fact that the Brownian motion is not sufficient to relax the structure formed under stress and that the suspension is able to store stress before yielding~\cite{Koumakis_2012, Zausch_2008}.\\
                                          
Moreover, at large deformation amplitudes, materials very often exhibit a response that depends on their deformation history. In a typical experiment, a pre-shear is applied, consisting in a constant or oscillatory deformation at high shear rate or high amplitude, and the mechanical property of the system is measured at some time $t_w$ after the end of this so-called rejuvenating protocol. For instance, the stress overshoot increases with the time elapsed since the rejuvenation, $t_w$. Past history dependence of the mechanical response of a system is a common feature that is well known in the engineering literature. Thus, metals exhibit a smaller resistance to compression when they have been previously submitted to tension. This effect has been discovered by Bauschinger when studying steels~\cite{Bauschinger_1881, Sowerby_1979}. A similar effect has been observed in atomistic simulations of amorphous systems \cite{Falk_1998, Karmakar_2010}, and discussed in the athermal limit, in terms of the elementary irreversible deformations that occur in such systems, in Shear Transformation Zone( STZ)~\cite{Falk_1998}. In that context, Patinet {\it{et al.}} ~\cite{Patinet_2020} have shown the importance of the role of the local stress barrier anisotropy to understand the origin of the Bauschinger effect in the deformation of athermal glasses.\\

More generally, the Bauschinger effect is related to hysteresis behavior of the stress when the applied deformation is cyclic and depends on the nature of the materials. It may, moreover, present some other specificities, such as a drastic reduction of the elastic response after the first cycle in filled elastomers (the so-called Mullins effect). As far as concentrated suspensions are concerned, they exhibit a reduction of their viscosity after the application of a high shear rate, a phenomenon called thixotropy. The behavior may be seen as a hysteresis response if one considers the stress \textit{vs} shear rate behavior :  the flow curve is measured upon increasing the applied shear rate and then upon decreasing the shear rate. These two flow curves do not superimpose and the area enclosed by the hysteresis loops decreases when the applied shear rate decreases. Experiments in carbopol suspensions~\cite{Divoux_2013} have shown that this behavior is also related to the appearance of shear banding.\\
 
Here, we study experimentally the nonlinear flow behavior of a concentrated suspension of spheres and which mimic hard sphere behavior reasonably well. We consider mainly the two behaviors described above :  stress overshoot and non-reversibility of the flow, that manifests itself by a hysteresis behavior and a change of the apparent elastic modulus upon flow reversal.  Moreover, we use numerical simulations to compute the anisotropy of the structure of the suspension submitted to the same deformation histories and  we study how the response is related to the local microstructure. 

\section{System}
\subsection{Experimental system}

We use spherical particles of cross-linked polymer chains, called microgels, that are swollen in a good solvent. Several such systems have been developed, in particular N-isopropylacrylamide particles cross-linked with  N,N'-methylenbisacrylamide, whose swelling ratio in water may be controlled by the temperature~\cite{Schild_1992, Senff_1999}. In the present study, non-thermosensitive
PS microgel particles are used~\cite{Beyer_2015, Eckert_2004}. Their structure and dynamical properties~\cite{Eckert_2003, Kegel_2000} have been studied in detail  and closely match that of hard sphere suspensions~\cite{Schneider_2017}. They have been used to study glass transition behavior~\cite{Eckert_2002} and crystallization kinetics~\cite{Palberg_2009, Iacopini_2009, Franke_2014}. We use bidisperse suspensions of particules with crosslinl ratio $1:50$ (\textit{i.e.} there is on average 1 crosslink per 50  monomer units) and  radii of $175$ nm and $150$ nm with number ratio $N=3.1$ ($3.1$ small particles \textit{per} $1$ large particle). The system exhibits a glass transition at $\phi=0.58$~\cite{Burger_2014}. 

\subsection{Rheological measurements}

An Anton-Paar stress controlled rheometer MCR-301 is used in order to perform rheological measurements. All measurements are done in a cone/plate geometry of radius $12.5$ mm and angle $2^\circ$. In a typical experiment, the sample is first rejuvenated by the application of an oscillatory deformation of  amplitude $\gamma = 400 \%$ and frequency $1$ Hz, for $300$ s. Then, the system is let at rest for $t_w=600$ s and the measurements are started. 
During all the experiments we follow a deformation protocol with a quasi-constant absolute value of the shear rate. In order to avoid inertia effects when the direction of flow is reversed, we apply a deformation protocol that follows a series of steps. The strain is incremented or decremented by  a constant value, $\delta \gamma= 5\cdot 10^{-4}$, separated by a constant time $\delta t$. The time of the deformation increase is negligibly small in comparison with $\delta t$ which is itself much smaller than the typical relaxation time of the system under study. We then define the effective shear rate as  $\dot{\gamma} = \delta\gamma/\delta t$ whose absolute value is kept constant during an experiment. The maximal deformation amplitude $\gamma_{max}$ is varied between $0.03$ and $0.2$. 

\subsection{Simulations}

In order to get the suspension microstructure, we perform molecular dynamic (MD) simulations using LAMMPS. We consider a binary mixture of particles, denoted $A$ and $B$, that interact through an inverse power law (IPL), potential. Experimenal measurements of the evolution of the elastic modulus with the volume fraction allows to infer the interaction potential between the particles, that is well described by~\cite{Schneider_2017}~: 
\begin{equation}
U_{\alpha\beta}(r_{ij}) = k_BT\left(\frac{\sigma_{\alpha\beta}}{r_{ij}}\right)^{42}
\end{equation}
\noindent where ${\alpha, \beta} ={A,B}$ and $r_{ij}$ s the distance between two particles. We choose $\sigma_{AA} =1$ and $\sigma_{BB} =0.876$. Assuming that the interactions are additive, the cross characteristic length scale is equal to $\sigma_{AB} = 0.938$. 
Particles are subjected to an overdamped Langevin dynamics at $T = 1.0$ with a  damping parameter $\xi= 4.0$. The time step is fixed to $\delta t = 10^{-3}$. The shear rate value is determined so that the experimental and numerical P\'eclet numbers are identical. We have~:

\begin{equation}
Pe = \dot{\gamma}\tau_B = \dot{\gamma}\frac{R^2}{D} = \dot{\gamma}\frac{R\xi}{k_BT}
\label{eq:Peclet}
\end{equation}

The experimental P\'eclet number is equal to $3\cdot 10^{-3}$, and the numerical shear rate is chosen in order to match the P\'eclet number obtained with Eq.~\ref{eq:Peclet}, taking into account the largest particles, of radius $R_{AA}=0.5$. This leads to the numerical shear rate value~: $\dot{\gamma} = 1.5\cdot 10^{-3}$. 
Simple shear deformation is obtained by the application of a succession of strain increments, $\delta \gamma$ during times $\delta t$, to the cubic box of size $L$. At each time step, the positions of the particles are updated according to~: $x'=x+\delta\gamma y$, $y'=y$ and $z'=z$. Then, the system is allowed to relax through viscous dissipation during time interval $\Delta t$. The applied shear rate is defined as $\dot{\gamma} = \frac{\delta \gamma}{\Delta t}$. \\
Cyclic shear experiments are performed, during which the simulation box is strained up to a value $\gamma=\gamma_{max}$. Then the shear is reversed, \textit{i.e} the strain increment is $-\delta \gamma$ and the system is deformed up to $\gamma =-\gamma_{max}$. Finally the cycle is completed by the application of shear in the positive direction back 
to $\gamma = 0$.\\

To compare the numerical simulations with the experiments, we have to consider equivalent packing fractions. To this aim, we use predictions of the mode coupling theory to estimate the packing fraction $\phi_c = 0.585$ at which the breakdown of ergodicity occurs. We then determine the relevant packing fractions $\phi$ from the separation parameter $\epsilon = \frac{\phi-\phi_c}{\phi_c}$. As a result, we numerically consider $\phi \in \{0.607, 0.617, 0.627\}$, corresponding to the  experimental volume fractions $\{0.60, 0.61, 0.62\}$. Throughout this work, we consider for each packing fraction $50$ samples of size $L = 60$.

To characterize the anisotropy of the suspension, we consider the fabric tensor $F$ which has been used successfully for instance in granular media and in silica or metallic glasses to track residual anisotropy during cyclic shearing. Following~\cite{Rountree_2009}, we define the local fabric tensor as~:
\begin{equation}
\overline{\overline{F}}(i) =\frac{1}{N_{nn}}\sum_{i,j\in n.n}\frac{\mathbf{r}_{ij}}{r_{ij}}\otimes\frac{\mathbf{r}_{ij}}{r_{ij}}
\end{equation}
\noindent where $\sum_{i,j\in n.n}$ denotes the sum taken over the nearest neighbors of particle $i$ and $N_{nn}$ is the number of nearest neighbors. The global fabric tensor is consequently expressed as
\begin{equation}
\overline{\overline{F}}= \frac{1}{N} \sum_{i=1}^N \overline{\overline F}(i)
\end{equation}

The global fabric tensor $\overline{\overline{F}} $ can be diagonalized and its eigenvalues $\lambda_1$, $\lambda_2$ and $\lambda_3$ measure the degree of isotropy in the $x,y,z$ directions. For a fully isotropic system one expects $\lambda_1=\lambda_2=\lambda_3=\frac{1}{3}$. We define the scalar quantity $\alpha$ which quantifies the degree of anisotropy~\cite{Rountree_2009}~:
\begin{equation}
\alpha = \frac{3}{2}\sqrt{\sum_{k=1}^3 \left(\lambda_k-\frac{1}{3}\right)^2}
\end{equation}

For a fully isotropic (\textit{resp.} anisotropic) system, $\alpha =0$ (\textit{resp.} $ \alpha =1$).

\subsection{The Generalized Maxwell Model}

We use the Generalized Maxwell Model, proposed by Voigtmann \textit{et al}~\cite{Voigtmann_2020, Frahsa_2013} in order to fit our experimental data. This model is based on the Integration Through Transients~\cite{Fuchs_2002} description of the flow of glassy systems and allows for an analytic expression of the $\sigma(\gamma)$ flow curve. 
 It is first assumed that the stress may be expressed as an integral over the past applied shear rates, that involves a memory function, $G(t,t')$~:
 \begin{equation}
 \sigma(t) =\int_{-\infty}^t \dot{\gamma}(t') G(t,t')
  \end{equation}
 \noindent where $G(t,t')$ depends on the applied deformation history.  It is written as~:
 
 \begin{equation}
 \nu_0\left(1-\left(\frac{\gamma(t,t')}{\gamma^*}\right)^2\right) e^{-\left(\frac{\gamma(t,t')}{\gamma^{**}}\right)^2}e^{-\frac{(t-t')\vert\dot{\gamma}\vert}{\gamma_c}}
 \end{equation}
 The model thus involves four parameters. $\nu_0$ is the elastic modulus at start-up flow when $\gamma\longrightarrow 0$, $\gamma_c/\vert \dot{\gamma}\vert$ is the characteristic shear induced relaxation rate value, and $\gamma^*$ and $\gamma^{**}$ capture the distribution of the shear induced relaxation rates over different length scales. In particular, it should be noted that this expression of the memory function allows $G(t,t')$ to take negative values, depending on the relative values of $\gamma^*$ and $\gamma^{**}$.

This model has been shown to exhibit a well-defined stress overshoot, and yields a hysteresis behavior when the deformation is reversed. 

\section{Startup flow and elasto-plastic transition}

\begin{figure}
\includegraphics[width = \textwidth]{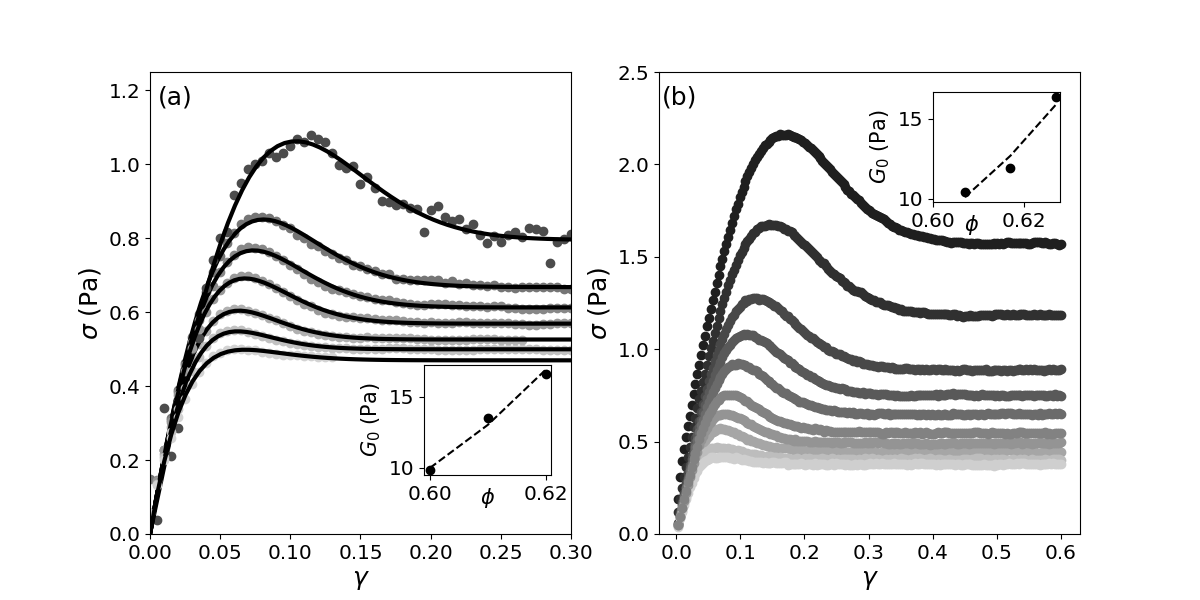}
\caption{\textbf{(a)}  Evolution of the stress as a function of the deformation, during the start-up flow. \textbf{(a)} Experimental data for $\phi=0.62$ and for different shear rate values, increasing from bottom. to top~: $\dot{\gamma} = 2.5\cdot 10^{-4},  5\cdot 10^{-4}, 10^{-3}, 2.5\cdot 10^{-3}, 5\cdot 10^{-3}, 10^{-2}, 5\cdot 10^{-2}$ s$^{-1}$. Continuous curves are adjustment with the Generalized Maxwell Model. \textbf{b)} Numerical data for $\phi=0.627$ and $\dot{\gamma}= 3\cdot 10^{-4}, 6\cdot 10^{-4}, 1.5\cdot 10^{-3}, 3\cdot 10^{-3}, 6\cdot 10^{-3}, 1.5\cdot 10^{-2}, 3\cdot 10^{-2}, 6\cdot 10^{-2},  1.5\cdot 10^{-1}$ and $3\cdot 10^{-1}$ s$^{-1}$. \textbf{Inset} of \textbf{(a)}  and \textbf{(b)} Evolution of the elastic modulus as a function of the volume fraction. The dashed lines are power-law adjustments, leading to $G_0\propto\phi^m$, with $m =14$ for the experimental data and $m=16$ for the numerical data.}
\label{fig:startupflow}
\end{figure}

Let us consider the start-up flow curve of suspensions of volume fractions $0.60$, $0.61$ and $0.62$ under the application  of a constant shear rate (Fig.~\ref{fig:startupflow} \textbf{(a)}). At low deformations, the response is linear and the elastic modulus is obtained and plotted in the inset of Fig.~\ref{fig:startupflow} \textbf{(a)} . The elastic modulus $G'$ at low deformation may be compared to the modulus obtained for suspensions of PMMA particles. We have, at $\phi=0.60$, $G'_0 = 9.8$ Pa $ = 11.6\frac{k_BT}{\langle R\rangle^3}$ Pa, close to the value $7.5\frac{k_BT}{R^3}$ obtained for PMMA particules suspensions of identical concentration~\cite{Amann_2013}. Moreover, we have $G'\propto \phi^m$ with $m=14$ (Fig.~\ref{fig:startupflow} \textbf{(a)} \textit{inset}), in agreement with measurements of Schneider~\cite{Schneider_2017} for similar system, where $m=14-17$ was reported.

\begin{figure}
\includegraphics[width = \textwidth]{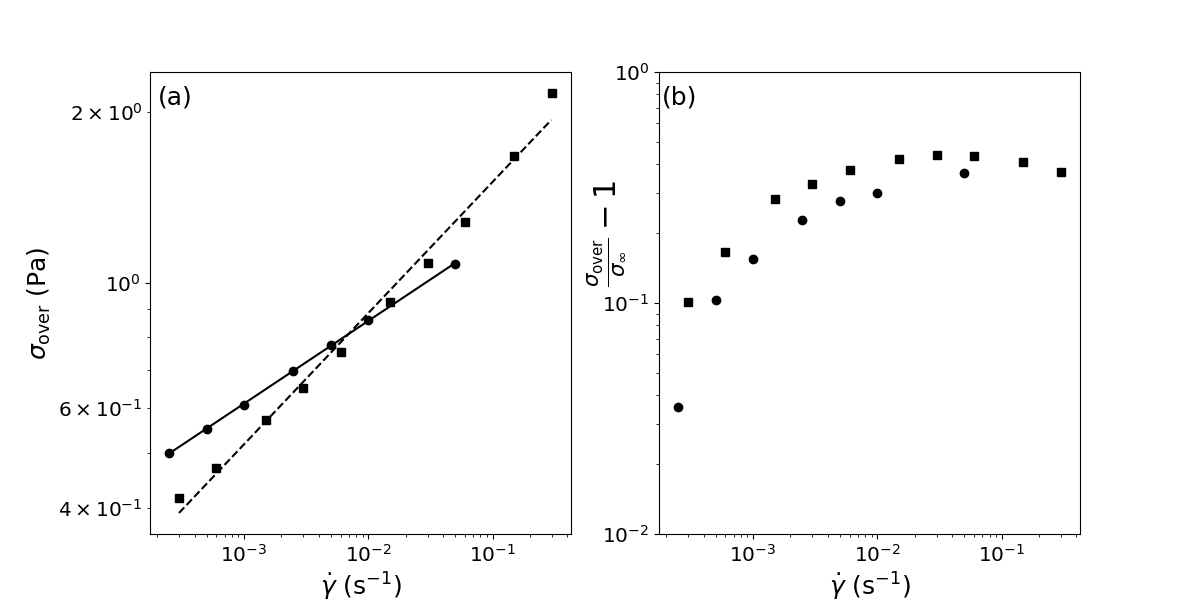}
\caption{ Evolution of the stress maximum at the overshoot \textbf{(a)} and of the overshoot stress relative to the stress at high shear rate \textbf{(b)} as function of the applied shear rate for  $\phi = 0.62$ (experimental data, $\bullet$) and for $\phi=0.627$ (numerical data, $\blacksquare$). \textbf{(a)} The lines are power law adjustment of the data, leading to $\sigma_{over}\propto\dot{\gamma}^\alpha$ with $\alpha = 0.14$ (continuous line, experimental data) and $0.16$ (dashed line, numerical data).}
\label{fig:overshoot}
\end{figure}

Then after yielding, characterized by stress overshoot, a plateau value of the stress is reached, corresponding to the flowing state. The numerical results obtained for the corresponding conditions are also given in Fig.~\ref{fig:startupflow} \textbf{(b)}. The evolution of the maximum stress value, $\sigma_{over}$ both from experiments and numerical simulations are given in Fig.~\ref{fig:overshoot} \textbf{(a)}. The amplitude of the stress overshoot increases when the shear rate increases~: during a given amount of time, more stress is relaxed under a lower shear rate. The overall set of flow curves may be adjusted by the Generalized Maxwell Model. In particular the memory function exhibits negative values, that allow for the description of stress overshoot. The amplitude and the position of this negative value is controlled by the parameters $\gamma^*$ and $\gamma^{**}$. These parameters are slightly varied between adjustment curves at different shear rates and are respectively equal to $0.074\pm 0.013$ and $0.092\pm 0.012$.\\
The evolution of the stress overshoot amplitude as a function of the shear rate is well described by $\sigma_{over}\propto\dot{\gamma}^\alpha$ with $\alpha =0.14$ for the experimental data and $\alpha = 0.16$ for the simulations. This value is close to the exponent obtained for attractive carbopol gel particules where shear banding is observed~\cite{Divoux_2011}. Although we do not have access to experimental measures of the local particle velocity, the numerical simulations do not exhibit any signature of shear banding, and the computed shear rate is constant through the gap of the cell.
One may also consider the amount of excess stress stored at yielding, relative to the stress corresponding to the maximal sheared state, under plastic flow, that is $\delta \sigma_{excess}  = (\sigma_{over} -\sigma_{\infty})/\sigma_{\infty}$. Experimental and numerical results are given in Fig.~\ref{fig:overshoot} \textbf{(b)}. This measures the ability of the system to store stress relative to the stress value of the suspension under a fully developed flow. At low shear rates, $\delta \sigma_{excess}$ increases with the shear rate, reflecting the fact that  the suspension is able to store more stress, but at higher shear rates the suspension cannot deform more than in the fully developed flow and the stress excess saturates. Our data are consistent with the results obtained in~\cite{Laurati_2012}, where the evolution of the stress also exhibits a saturation of excess of stress at high shear rates.
                               
\section{Flow reversal behavior}

\begin{figure}
\includegraphics[width = \textwidth]{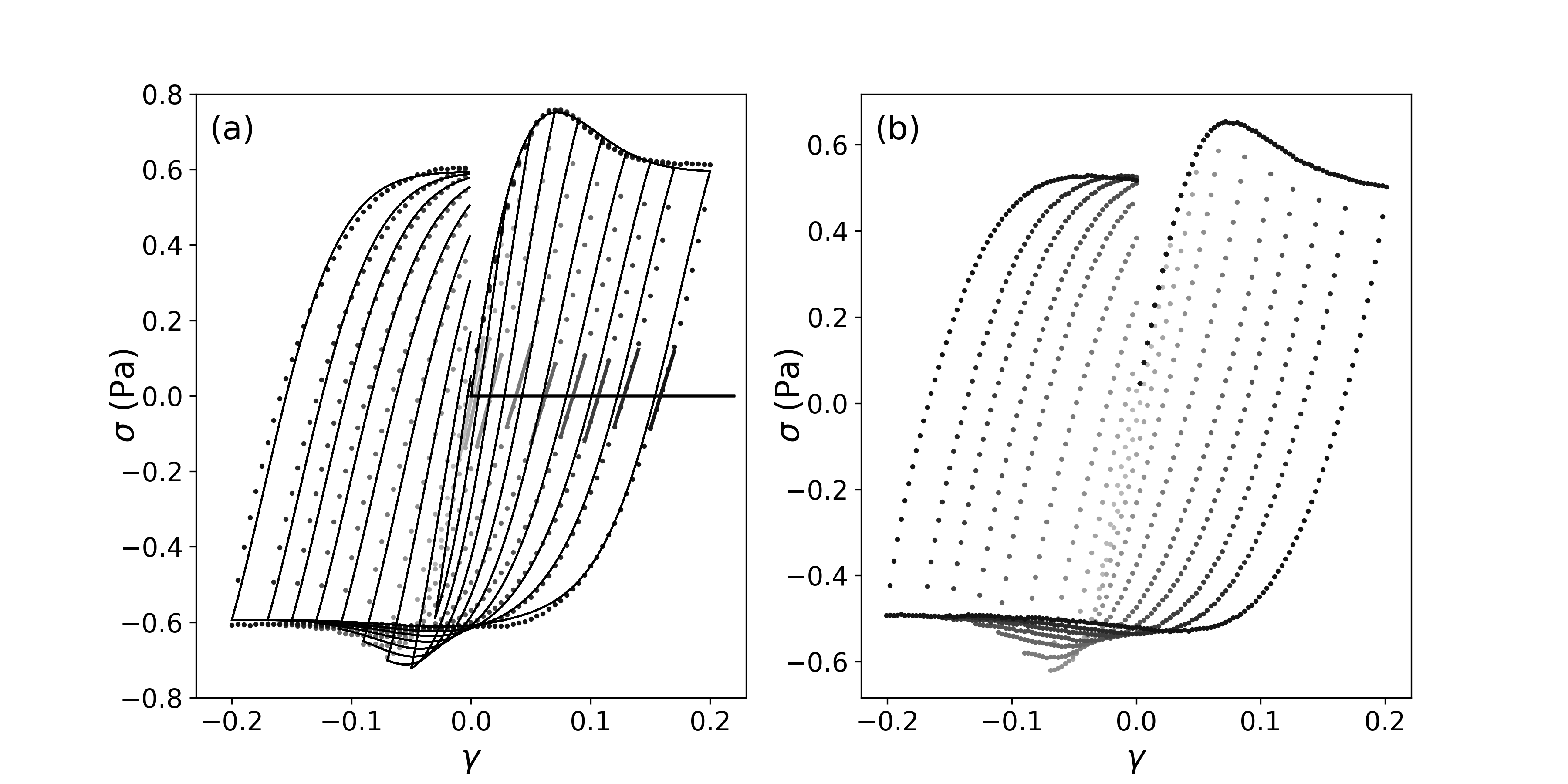}
\caption{Stress as a function of the deformation during the first cycle of deformation : the deformation is increased from $0$ to $\gamma_{max}$; then decreased from $\gamma_{max}$ to $-\gamma_{max}$ and increased back from $-\gamma_{max}$ to $0$. at a constant absolute value of the shear rate. \textbf{(a)} Experimental data for $\phi=0.62$ and $\gamma_{max}= 0.03$, $0.05$, $0.07$, $0.09$, $0.11$, $0.13$, $0.15$, $0.17$ and $0. 20$. The continuous black lines are adjustment of the data with the Generalized Maxwell Model using a single set of parameter values for all cycles. The thick segments are linear adjustment of the $\sigma(\gamma)$ curves around zero stress, after flow reversal, from with the effective modulus, $G_{back}$ is computed.  \textbf{(b)} Numerical data for $\phi=0.627$ and for the corresponding set of $\gamma_{max}$ values as experimental data.}
\label{fig:hysteresis}
\end{figure}

Let us now consider the flow behavior after flow reversal. The suspension is first submitted to a constant shear rate $\dot{\gamma}= 5 \cdot 10^{-3}$ s$^{-1}$ up to a deformation value $\gamma_{max}$ and then to a reverse flow of same shear rate.  The evolution of the stress is given in Fig.~\ref{fig:hysteresis}, for $\phi=0.62$, both experimental results \textbf{(a)} and numerical results\textbf{(b)} are given.  A strong hysteresis behavior is observed, already reported in athermal systems~\cite{Fiocco_2013}. The stress/strain relationship after flow reversal is markedly different from the startup flow curve. One gets a strong reduction of the stress overshoot. Adjustment of the reverse flow curves may be obtained from the GMM model using the parameters determined from the adjustment of the startup flow curve, and reproduce the reversal flow behavior. In order to understand this reduction, we consider the memory function $G$ of the GMM model. As already discussed, during the startup flow, the memory function exhibits negative values responsible for the stress overshoot; upon flow reversal, the memory is erased by the $e^{-\gamma/\gamma_c}$ factor and the overshoot is reduced. 

\begin{figure}
\includegraphics[width = \textwidth]{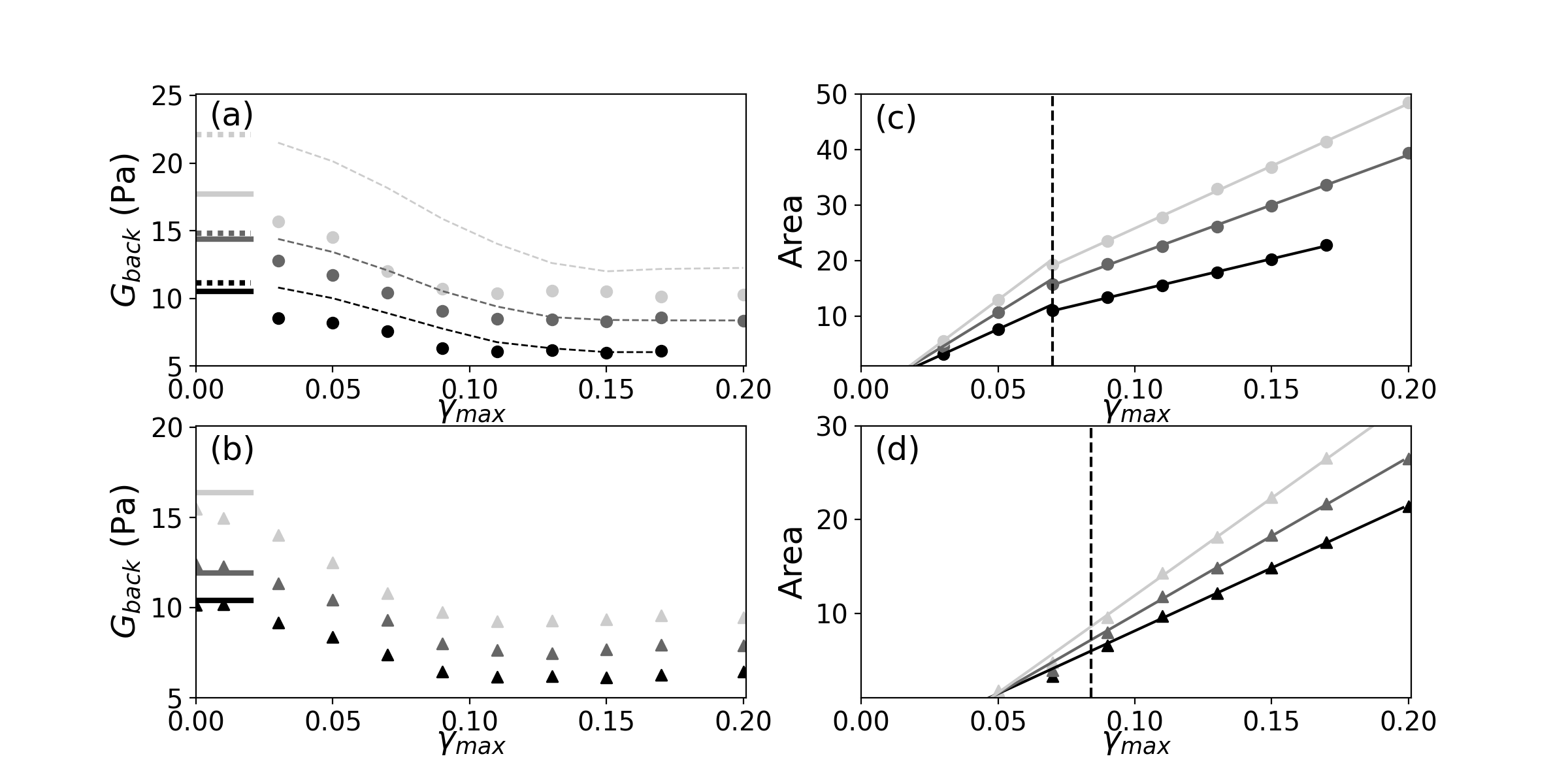}
\caption{\textbf{(a)} and \textbf{(b)} Evolution of the effective modulus, measured as the slope of the $\sigma$ \textit{vs} $\gamma$ curve when the stress is null, after flow reversal (see Fig.~\ref{fig:hysteresis}, for experimental systems \textbf{(a)} and numerical systems \textbf{(b)}, as a function of the maximum deformation applied, $\gamma_{max}$.  In \textbf{(a)} dashed lines are the evolutions of $G_{back}$ obtained from the adjustment of the experimental data with the Generalized Maxwell Model (see also Fig.~\ref{fig:hysteresis}). The horizontal thick lines correspond to the elastic modulus at rest, defined as the slope of the startup flow curves when $\gamma \longrightarrow 0$ (in \textbf{(a)}, continuous lines~: experimental data, dotted lines~: GMM data).
\textbf{(c)} and \textbf{d)} Evolution of the area of the hysteresis cycles as a function of their maximum amplitude $\gamma_{max}$. Vertical dashed line indicates the $\gamma$ value at the stress overshoot, $\gamma_{over}$. Continuous lines are linear adjustments of the data for $\gamma$ values larger and smaller than the deformation at stress overshoot. Three volume fractions are studied, increasing from black to grey : $\phi=0.60$, $0.61$ and $0.62$ for the experimental system and $\phi=0.607$, $0.617$ and $0.627$ for the numerical system. } 
\label{fig:Bauschinger}
\end{figure}

Experimental and numerical results exhibit a softening of the suspensions upon flow reversal (Fig.~\ref{fig:Bauschinger} \textbf{(a)} and \textbf{(b)}). We compute the effective shear modulus $G_{back}$ as the slope of the stress \textit{vs} strain curve when the total stress gets back to zero (the slope is plotted in Fig.~\ref{fig:hysteresis} \textbf{(a)}). This effective modulus upon shear reversal is smaller than the elastic modulus at rest and decreases when the maximum applied amplitude increases, reaching a plateau value at large $\gamma_{max}$. This trend is well reproduced by the Generalized Maxwell Model. Nevertheless, at the highest studied concentration, the amplitude of the decay is strongly overestimated by the model. More precisely, the GMM cannot describe quantitatively both the value of the overshoot and the amplitude of the reduction of the apparent elasticity after flow reversal and a slight overestimation is also visible for the other volume fractions. This indicates that the model does not capture the overall relaxation mechanisms that play a role in the non-linear response of the suspension, in particular at the largest volume fractions. \\
Another characteristic feature of the non-linear behavior of the response is its hysteretic behavior. The area enclosed by the hysteresis curves is represented in Fig.~\ref{fig:Bauschinger} \textbf{(c)} and \textbf{(d)}. At low maximal deformation amplitudes, the area enclosed by the hysteresis curves is close to $0$, and the response is close to purely elastic. Then, with increasing value of $\gamma_{max}$, the amplitude enclosed by the curves increases. Moreover the experimental evolution of the area as a function of  $\gamma_{max}$ exhibits two behaviors, corresponding to $\gamma_{max}$ larger or smaller than the deformation at which the overshoot is observed, $\gamma_{over}$~: at large maximum deformation values, the enclosed area increases slower than at smaller deformations, which may be understood as the fact that the stress reaches a constant value at high $\gamma_{max}$. This behavior is not recovered in simulations, where the slope of the area as a function of $\gamma_{max}$ does not decrease at $\gamma_{over}$. This is due a slight difference in the width of the hysteretic response at small deformation amplitudes. \\

These results show that, even though flow non-linearities are qualitatively explained by a simple model of the relaxation mechanisms and may be reproduced by numerical simulations, the exact dissipation mechanisms require a more precise description of the interaction potential between the particles and a more refined relaxation spectra. 

\section{Microscopic interpretation}
\begin{figure}
\includegraphics[width = \textwidth]{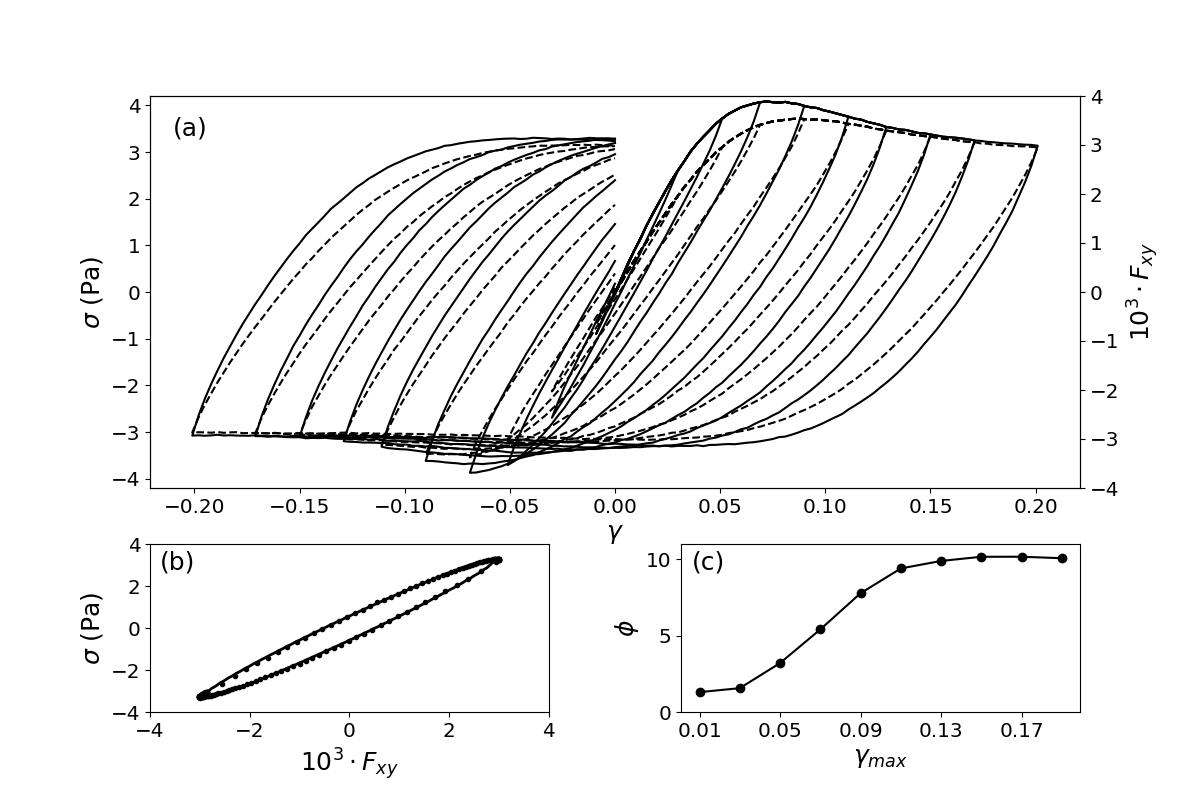}
\caption{\textbf{(a)} Evolution of the stress (left axis, continuous lines) and of the $xy$ component of the fabric tensor(right axis, dashed line) for a suspensions of volume fraction $0.627$ submitted to cycles of maximum amplitudes $\gamma_{max} = 0.03$, $0.05$, $0.07$, $0.09$, $0.11$, $0.13$, $0.15$, $0.17$ and $0.20$. \textbf{(b)} Stress as a function of the $xy$ component of the fabric tensor for $\gamma_{max}= 0.2$. The continuous line is an elliptic adjustment of the data obtained by Principal Component Analysis. \textbf{(c)} Evolution of the phase shift $\phi$ between the stress and  $F_{xy}$, deduced from the elliptical adjustement of the $\sigma(F_{xy})$ curve.}
\label{fig:microstructure}
\end{figure}

To understand the microscopic origin of the Bauschinger effect in the colloidal suspensions, we follow the anisotropy during the deformation and, in particular, we characterize the possible differences between states when no constraint is applied on the systems, i.e $\sigma = 0$.
As we perform simple shear in the $xy$ plane, we first look at the evolution of $F_{xy}$, the $xy$ component of the fabric tensors with strain. The hysteresis cycles of the stress and of the $xy$ component of the fabric tensor, $F_{xy}$ are plotted in Fig.~\ref{fig:microstructure} \textbf{(a)}, for a series of shear maximum amplitudes, $\gamma_{max}$. $F_{xy}$ follows a similar hysteretic cycle as the stress : at low amplitudes, the anisotropy increases linearly with the deformation and takes the same values when the shear is reversed. When the amplitude increases, stress overshoot develops, and corresponds to an overshoot of the anisotropy of the microstructure. Nevertheless, the evolution of $F_{xy}$ as a function of the deformation is delayed, when compared to the evolution of the stress. In particular, after flow reversal, whereas the overall stress comes back to a null value, the microscopic structure still exhibits an
anisotropy characterized by a positive value of $F_{xy}$. The suspension has not relaxed back to its equilibrium isotropic structure and it might be expected that a different linear response is obtained. By plotting the evolution of $F_{xy}$ with the stress (Fig.~\ref{fig:microstructure} \textbf{(b)}), one may quantify the delay of the microstructure anisotropy relative to the stress. The retardation of the anisotropy of the microstructure may be quantified by the ellipticity value obtained by singular value decomposition of the data. The results are given in  Fig.~\ref{fig:microstructure} \textbf{(c)}~: the phase lag between the stress and the anisotropy is plotted as a function of the maximum applied deformation. It increases during the initial startup flow phase, until the deformation corresponding to the stress overshoot and then remains constant as a function of the maximum applied deformation. This corresponds to the observation of the decay of the effective modulus, that decays at low maximum shear amplitudes and reaches a plateau at $\gamma \approx 0.1$. This indicates that the Bauschinger effect is associated to a phase lag between the stress and the deformation that builds up during the initial startup flow and remains constant when under the established plastic flow regime.

\section{Conclusion}

Concentrated colloidal suspensions exhibit a rich flow  behavior, and their mechanical properties depend on the previously applied stress. 
In particular, after flow reversal, the deformation is not reversible. This leads to a hysteresis curve in the $\sigma(\gamma)$ plane whose area  is the energy dissipated during a cycle. Moreover, the shape of the hysteresis curve does not possess a central symmetry relative to the non deformed state at zero stress, and shearing in an opposite direction leads to a different response from the deformation in the direction from a rest state~: the stress overshoot disappears and the apparent elastic modulus is reduced after flow reversal. These observations are in agreement with numerical simulations~\cite{Frahsa_2013} and our own simulations show that the anisotropy of the suspension microstructure is retarded relative to the macroscopic stress after flow reversal.

\newpage

\newpage

\newpage

\newpage

\newpage

\clearpage
%
\end{document}